\documentclass[usenatbib]{mn2e}
\include{graphicx}

\def\deg{\hbox{$^\circ$}}

\def\msun{M_\odot}
\def\rsun{R_\odot}

\def\namefull{2MASS~J01542930+0053266}
\def\name{J0154}
\def\et{et~al.\ }

\def\myeps@scaling{.55}
\def\myplotone#1{\centering \leavevmode
    \epsfxsize=\myeps@scaling\columnwidth \epsfbox{#1}}

\begin{document}
\title[\namefull]{\namefull : A New Eclipsing M--dwarf Binary System}
\author[A.C.~Becker \et]{A.C.~Becker$^{1}$, % becker@astro.washington.edu
E.~Agol$^{1}$, % agol@astro.washington.edu
N.M.~Silvestri$^{1}$, % nms@astro.washington.edu
J.J.~Bochanski$^{1}$, % bochansk@astro.washington.edu
C.~Laws$^{1}$, % laws@astro.washington.edu
\newauthor
A.A.~West$^{1,2}$, % awest@astro.berkeley.edu\
G.~Basri$^{2}$, % basri@berkeley.edu
V.~Belokurov$^{3}$, % vasily@ast.cam.ac.uk
D.M.~Bramich$^{3,4}$, % dmb7@ast.cam.ac.uk
J.M.~Carpenter$^{5}$, % jmc@astro.caltech.edu
\newauthor
P.~Challis$^{6}$, % pchallis@cfa.harvard.edu
K.R.~Covey$^{6}$, % kcovey@cfa.harvard.edu
R.M.~Cutri$^{7}$, % roc@ipac.caltech.edu
N.W.~Evans$^{3}$, % nwe@ast.cam.ac.uk
M.~Fellhauer$^{3}$, % madf@ast.cam.ac.uk
\newauthor
A.~Garg$^{6}$, % artigarg@fas.harvard.edu
G.~Gilmore$^{3}$, % 	gil@ast.cam.ac.uk
P.~Hewett$^{3}$, % phewett@ast.cam.ac.uk
P.~Plavchan$^{7}$, % plavchan@ipac.caltech.edu
D.P.~Schneider$^{9}$, % dps@xl5.astro.psu.edu
\newauthor
C.L.~Slesnick$^{5}$,% cls@astro.caltech.edu
S.~Vidrih$^{3,8}$, % simon.vidrih@fmf.uni-lj.si
L.M.~Walkowicz$^{1}$, % lucianne@astro.washington.edu
and D.B.~Zucker$^{3}$ % zucker@ast.cam.ac.uk
\\
%-------------------- institutions and email ----------------------
$^1$ Astronomy Department, University of Washington, Seattle, WA 98195 \\
$^2$ Astronomy Department, University of California, 601 Campbell Hall, Berkeley, CA 94720-3411 \\
$^3$ Institute of Astronomy, University of Cambridge, Madingly Road, Cambridge, CB3 0HA, UK \\
$^4$ Isaac Newton Group of Telescopes, Apartado de Correos 321, E-38700 Santa Cruz de la Palma, Canary Islands, Spain \\
$^5$ Department of Astronomy, California Institute of Technology, Mail Code 105-24, 1200 East California Boulevard, Pasadena, CA 91125 \\
$^6$ Physics Department, Harvard University, 17 Oxford Street, Cambridge, MA 02138 \\
$^7$ Infrared Processing and Analysis Center, M/C 100-22, 770 S. Wilson Ave, Caltech, Pasadena, CA 91125 \\
$^8$ Astronomisches Rechen-Institut/Zentrum f\"ur Astronomie der Universit\"at  Heidelberg, M\"onchhofstrasse 12-14, 69120 Heidelberg, Germany \\
$^9$ Department of Astronomy and Astrophysics, Pennsylvania State University, 525 Davey Laboratory, University Park, PA 16802, USA
}

\maketitle
\begin{abstract} 
  We report on \namefull, a faint eclipsing system composed of two M
  dwarfs.  The variability of this system was originally discovered
  during a pilot study of the 2MASS Calibration Point Source Working
  Database.  Additional photometry from the Sloan Digital Sky Survey
  yields an 8--passband lightcurve, from which we derive an orbital
  period of $2.6390157 \pm 0.0000016$ days.  Spectroscopic followup
  confirms our photometric classification of the system, which is
  likely composed of M0 and M1 dwarfs.  Radial velocity measurements
  allow us to derive the masses (M$_1 = 0.66 \pm 0.03 \msun$; M$_2 =
  0.62 \pm 0.03 \msun$) and radii (R$_1 = 0.64 \pm 0.08 \rsun$; R$_2 =
  0.61 \pm 0.09 \rsun$) of the components, which are consistent with
  empirical mass--radius relationships for low--mass stars in binary
  systems.  We perform Monte Carlo simulations of the lightcurves which
  allow us to uncover complicated degeneracies between the system
  parameters.  Both stars show evidence of H$\alpha$ emission,
  something not common in early--type M dwarfs.  This suggests that
  binarity may influence the magnetic activity properties of low-mass
  stars; activity in the binary may persist long after the dynamos in
  their isolated counterparts have decayed, yielding a new potential
  foreground of flaring activity for next generation variability
  surveys.

\end{abstract}

\section{Introduction}
\label{sec-intro}

Low--mass dwarfs ($0.07\msun \leq$ M$_{\star} \leq 0.7\msun$) comprise
$\sim$ 75\% of all stars in the Milky Way, making them the most common
luminous objects in the Galaxy \citep{1995AJ....110.1838R}.
%
%Measuring the fundamental parameters, radiative processes, and the
%interior physics of stars on the lower main sequence is difficult due
%to their diminutive size, relative faintness, and complex atmospheric
%molecular compositions.  A significant amount of theoretical work has
%been done to construct low-mass main sequence stellar models
%\citep[e.g.][]{1993ApJ...406..158B,1998A&A...337..403B,1999ApJ...512..377H,2000ARA&A..38..337C},
%but distinguishing between the various models is difficult without
%precise ($\sim 2\%$) empirical constraints
%\citep{2006Ap&SS.304...89R}.  Double--lined eclipsing binaries with
%detached, low--mass components of similar spectral type offer the best
%opportunity for accurate and precise measurements of these fundamental
%properties.
% COVEY
A significant amount of theoretical work has been devoted to
constructing models that describe the physical processes in the
interiors and atmospheres of these low--mass stars
\citep[e.g.][]{1993ApJ...406..158B,1998A&A...337..403B,1999ApJ...512..377H,2000ARA&A..38..337C}.
Differences in the predictions of these models can be subtle, and
distinguishing between them requires precise empirical constraints on
fundamental stellar properties \citep[mass, radius, luminosity, and
effective temperature;][]{2006Ap&SS.304...89R}.  As low--mass stars
are faint, small, and possess complex spectra dominated by strong
molecular bands, measuring these parameters is challenging.
Double--lined eclipsing binaries with detached, low--mass components
of similar spectral type offer the best opportunity for accurate and
precise measurements of these fundamental properties.

Although binary systems are more common than single stars at high
stellar masses \citep[$>$ 1 M$_{\odot}$; eg.][]{Abt-83,Duquennoy-91},
binaries are not very common in low--mass stars
\citep{1997A&A...325..159L,1997AJ....113.2246R,2004ASPC..318..166D}.
In fact, the combination of low binary fraction and the sheer
dominance (by number) of low--mass stars suggests that most of the
stars in the Galaxy are actually single stars
\citep{2006ApJ...640L..63L}.  Because of the low binary fractions and
the faintness of M dwarfs, relatively few low--mass main sequence
binaries have been studied \citep[e.g.][and references
therein]{Creevey-05,LopezM-05,Hebb-06,Ribas-03,2006Ap&SS.304...89R,Blake-07}.
Adding to the census of double--lined eclipsing binary systems is
critical because they provide highly accurate, direct measurements of
the masses and radii of the components, nearly independent of any model
assumptions.  While the known number of eclipsing low--mass binaries
is small, next-generation variability
% (LSST, Pan-STARRS, and GAIA) 
and planet hunting 
% (Kepler, TrES, WASP) 
surveys should increase the number of known systems.

Measurements of the masses and radii of the individual components of
these low--mass systems have revealed potentially serious inadequacies
in stellar evolution models \citep{Ribas-03,2006Ap&SS.304...89R},
whereas measurements and models agree for stars over $ 1\msun$.  These
discrepancies apparently extend down into the brown dwarf regime as
evidenced by the first L--dwarf binary \citep{Stassun-06,
2007ApJ...664.1154S}.  In particular, the mass-radius,
mass-temperature, and mass-luminosity relationships predicted by
stellar theory are inconsistent with a large fraction of observed M
dwarf components \citep[e.g.][]{Hebb-06}.  Identifying and
characterizing new low--mass eclipsing binaries will provide stronger
constraints for theoretical models and help reveal the cause of the
current discrepancies between observed and predicted properties of
low--mass stars.

In this manuscript, we present the discovery of a new double-lined
eclipsing binary system, \namefull.  In \S 2, we outline the
photometric and spectroscopic observations of this binary system.  Our
determination of physical parameters and spectral types are detailed
in \S 3.  Discussion of the discrepancies between empirical and
theoretical mass-radius relations is outlined in \S 4, followed by our
conclusions in \S 5.

\section{Observations}
\subsection{2MASS Photometry}
\label{sec-id}
The calibration observations of the Two Micron All Sky Survey
\citep[2MASS;][]{2006AJ....131.1163S} are a precursor to
next--generation survey efforts such as LSST
\citep{2002SPIE.4836...10T}.  2MASS observed the entire sky using
three--channel cameras that simultaneously imaged in the $J$ (1.25
$\mu$m), $H$ (1.65 $\mu$m), and $K_{\rm s}$ (2.17 $\mu$m) passbands.
To allow precision cross--calibration of the data, 35 $1\deg \times
8.\arcmin 5$ calibration fields were defined.  An available
calibration field was imaged approximately hourly over the duration of
the survey.  The total number of epochs obtained for a given
calibration field ranges from 600 to 3700.  The 2MASS Calibration
Point Source Working Database (Cal-PSWDB) has recently been released
as part of the 2MASS Extended Mission \citep{Cutri-2massExtended}, and
contains a wealth of information on temporal variability
\citep{plav-07,becker-cutri-08}.

\namefull\footnote{For completeness, we note that the averaged
  position of this object in the Calibration Point Source Working
  Database is 01542929+0053272.  Because the calibration source
  astrometry has a slight bias relative to the survey, having been
  derived from different astrometric standards, we use the designation
  of the object from the 2MASS All-Sky Point Source Catalog.}\
(hereafter \name) was first identified as a periodic variable in a
pilot study of objects in 2MASS calibration region 90004, which was
imaged 2977 times between July 1997 and November 2000.  We extracted
lightcurves for 5770 individual objects from the ensemble of data by
clustering measurements taken at different epochs based on their
celestial coordinates.  For a given lightcurve, we used the 2MASS
photometric quality flag $ph\_qual$ to reject poor--quality data
points ($ph\_qual$ = D, E, F, or X).
%Thus, most of our 2MASS photometric data has an uncertainty of $sim
%5\%$ \citep{Cutri-2massExtended}.  
In total, 965 of these clipped lightcurves had more than 100 epochs.
We phased all clipped lightcurves using a modified version of the
Supersmoother algorithm \citep{Riemann-94}.  For each object, the $J$,
$H$, and $K_{\rm s}$ passbands were phased independently, and the
best--fit periods were compared.  The phasing of \name\
%
%at coordinates $\alpha$ = 01:54:29.30 $\delta$ = +00:53:27.27 (J2000),
%
yielded periods of $1.31951246$ days in all three passbands (we note
that Supersmoother converged upon an alias of the final period derived
for this system).  The composite colours of the system ($J - H =
0.66$, $H - K = 0.19$) suggest it consists of early--type dM stars
\citep{2007arXiv0707.4473C}.

\subsection{SDSS Photometry}
\label{subsec-obs:sdss}
\name\ lies in the Sloan Digital Sky Survey's
\citep[SDSS;][]{1998AJ....116.3040G,2000AJ....120.1579Y,2002AJ....123..485S,2003AJ....125.1559P,2006AJ....131.2332G}
Stripe 82, a $\sim$ 300 sq. deg. equatorial region that has been
imaged repeatedly over the course of the SDSS.  Stripe 82 is imaged by
the SDSS--II Supernova Survey every other night from September to
December, 2005--2007 \citep{2007arXiv0708.2749F}.  The extensive
repeat imaging of this region of sky has enabled precise photometric
and astrometric calibration of this Stripe, yielding the Stripe 82
Light-Motion Curve Catalogue \cite[LMCC;][]{Bramich-LMCC}.  We
extracted the light--motion curve for \name\ from the LMCC, which
consists of 32 $u,g,r,i$ and $z$--band
\citep{1996AJ....111.1748F,2002AJ....123.2121S} measurements of the
system, including SDSS--I observations as far back as Sept, 1998 and
up to the end of SDSS--II supernova survey observations in Dec, 2005.
The catalogue reports a proper motion vector for the system of
$\mu_{\alpha}$ = $0.88 \pm 2.25$ mas yr$^{-1}$ and $\mu_{\delta} =
-11.19 \pm 2.25$ mas yr$^{-1}$.

\begin{figure*}
  \includegraphics[width=8cm]{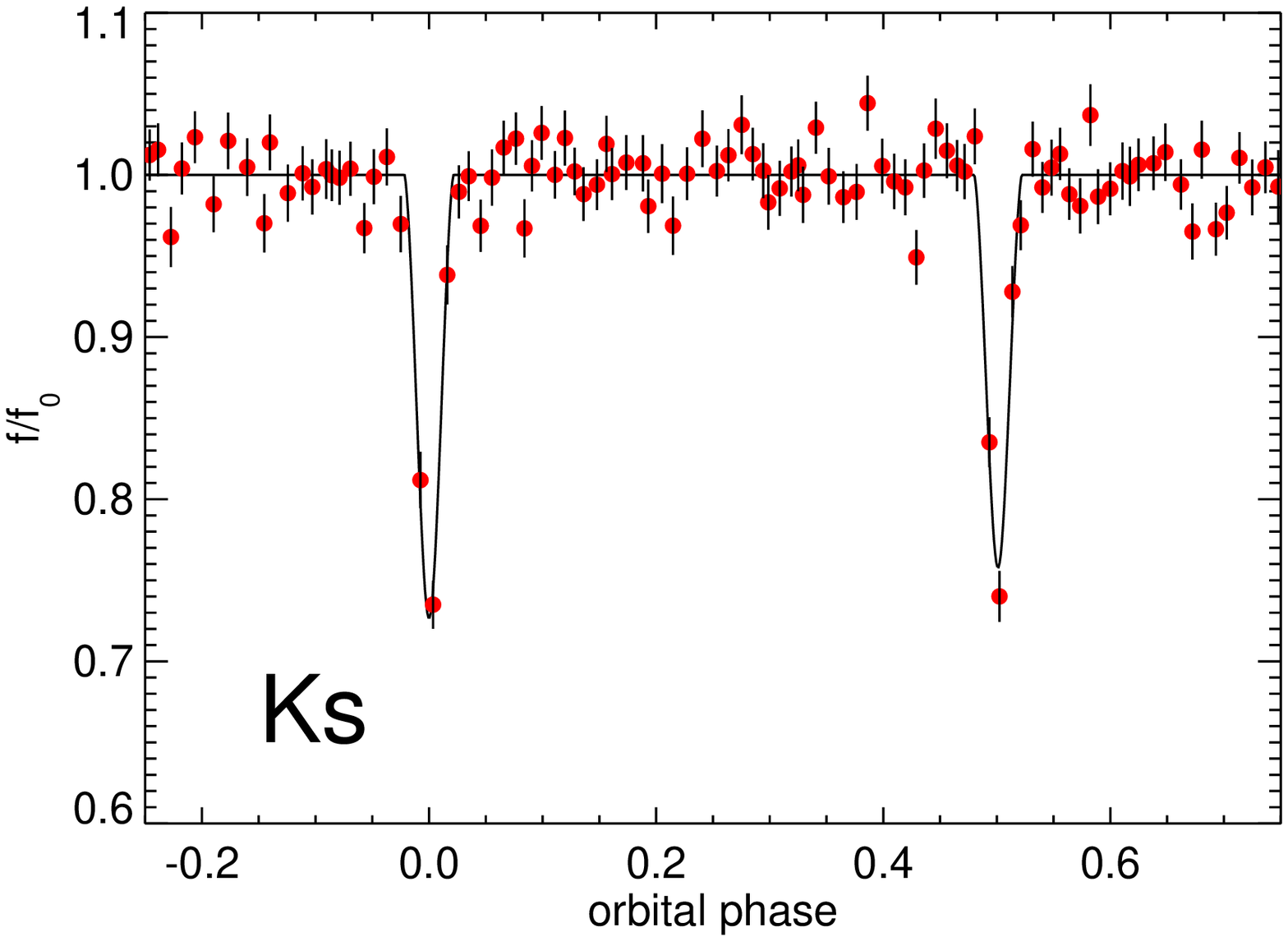}
  \includegraphics[width=8cm]{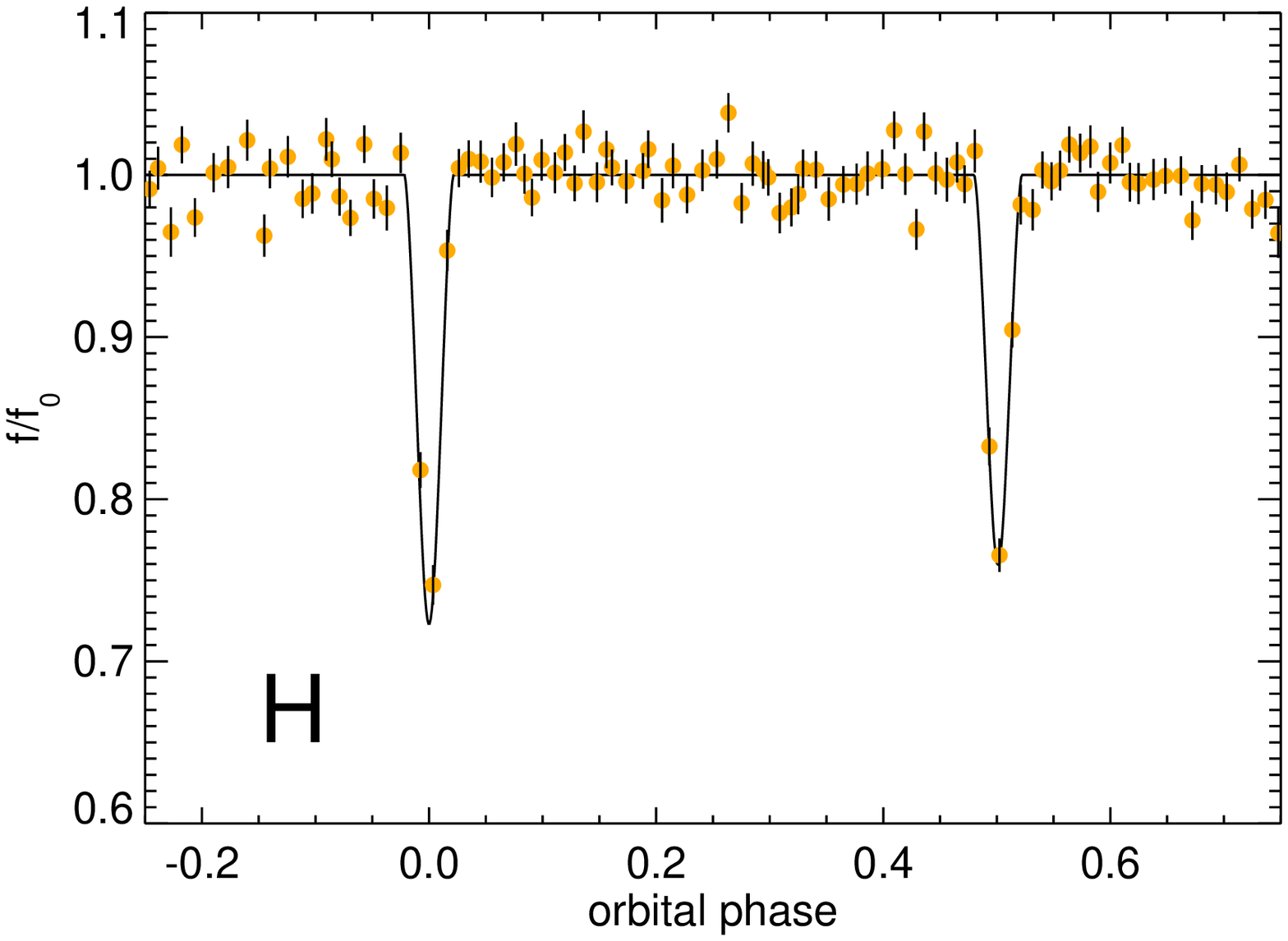}
  \includegraphics[width=8cm]{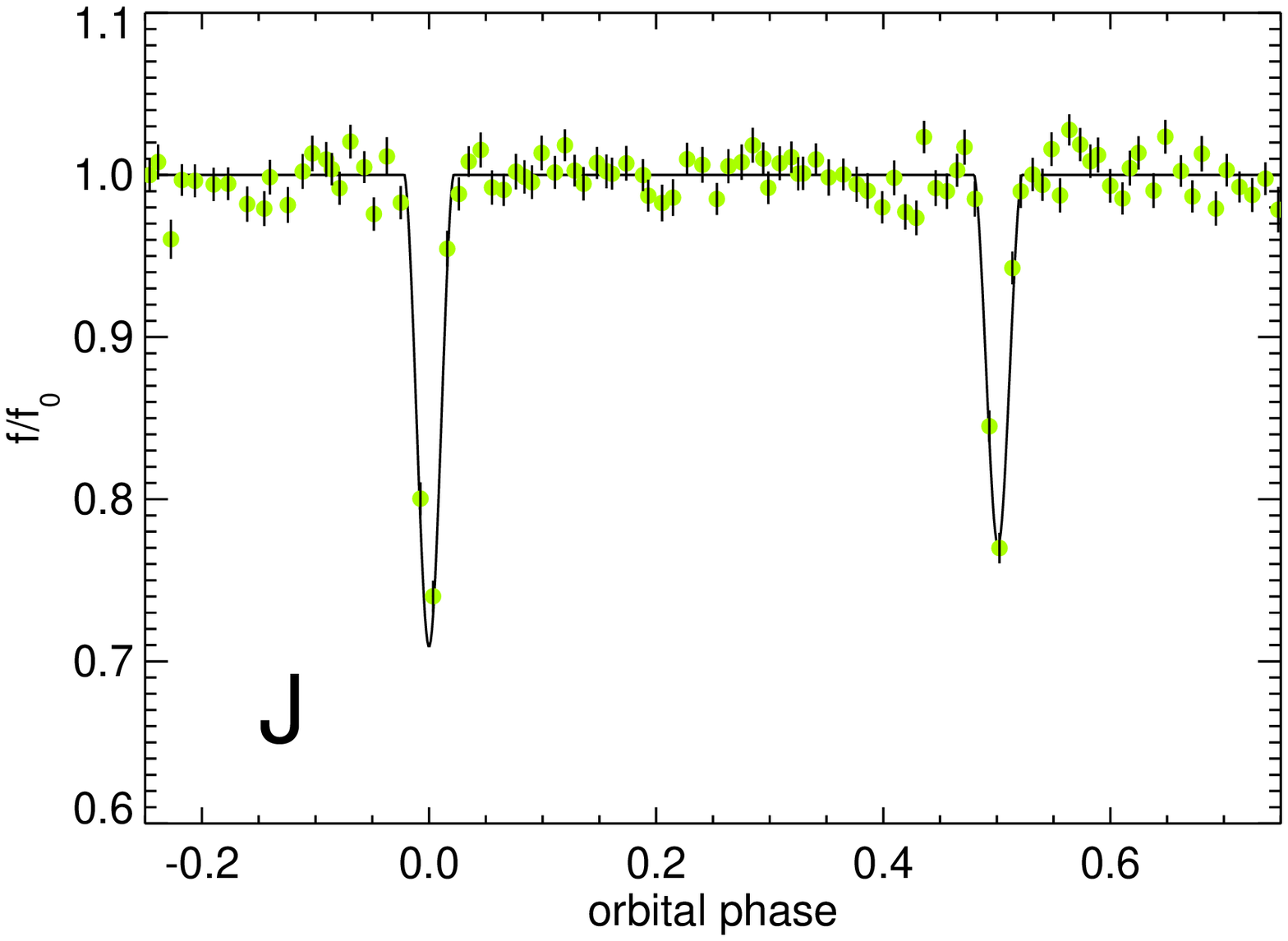}
  \includegraphics[width=8cm]{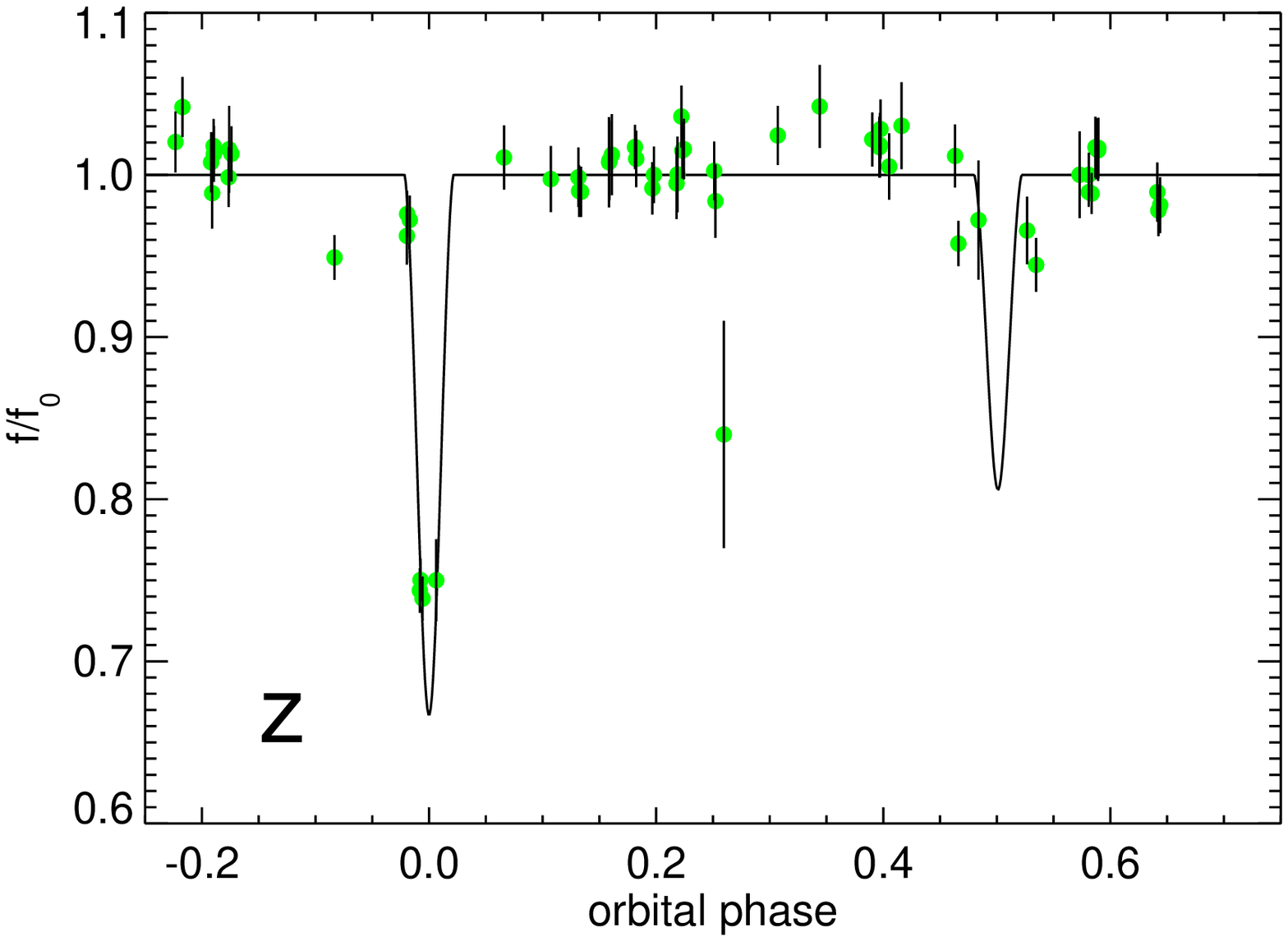}
  \includegraphics[width=8cm]{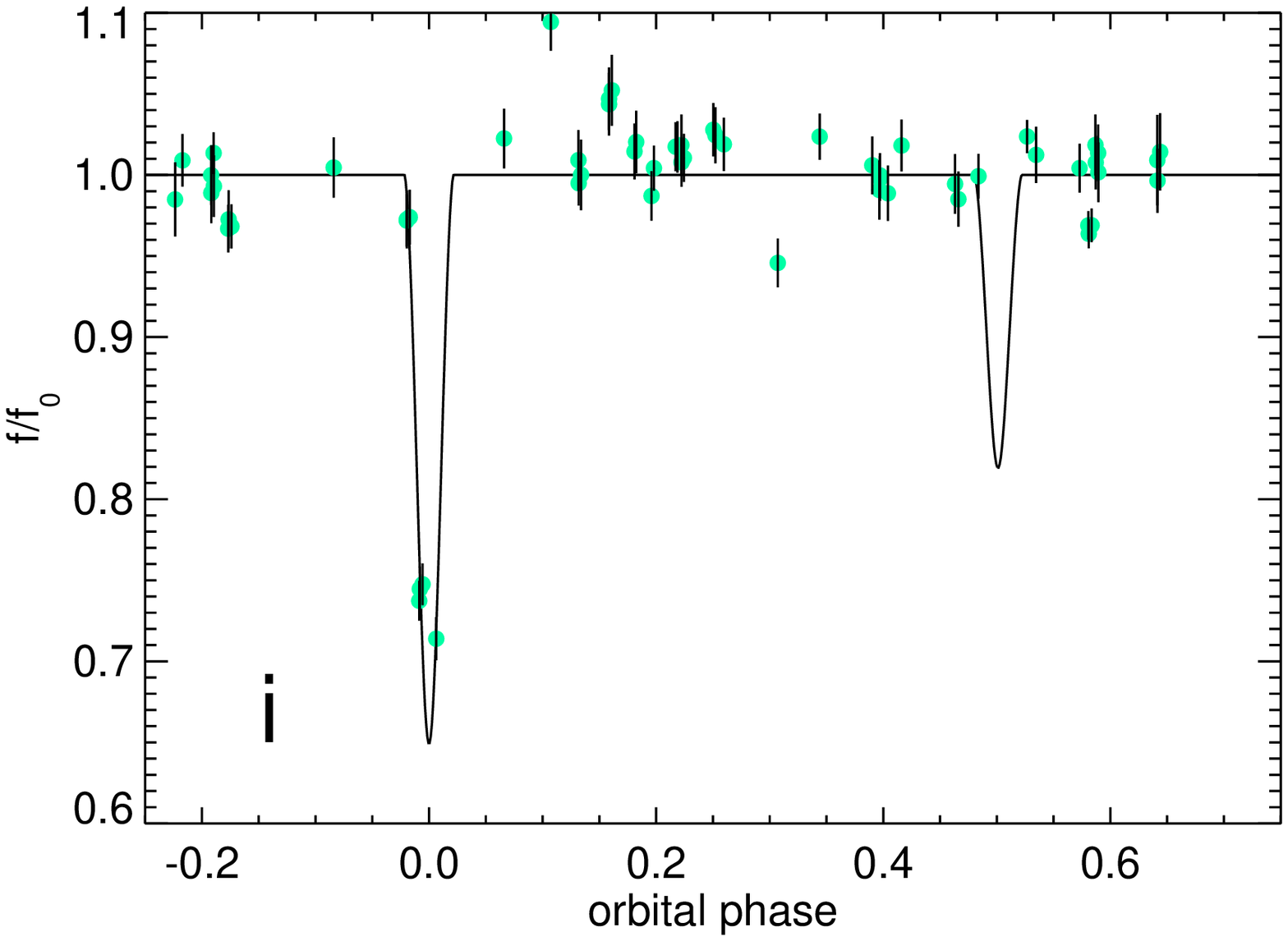}
  \includegraphics[width=8cm]{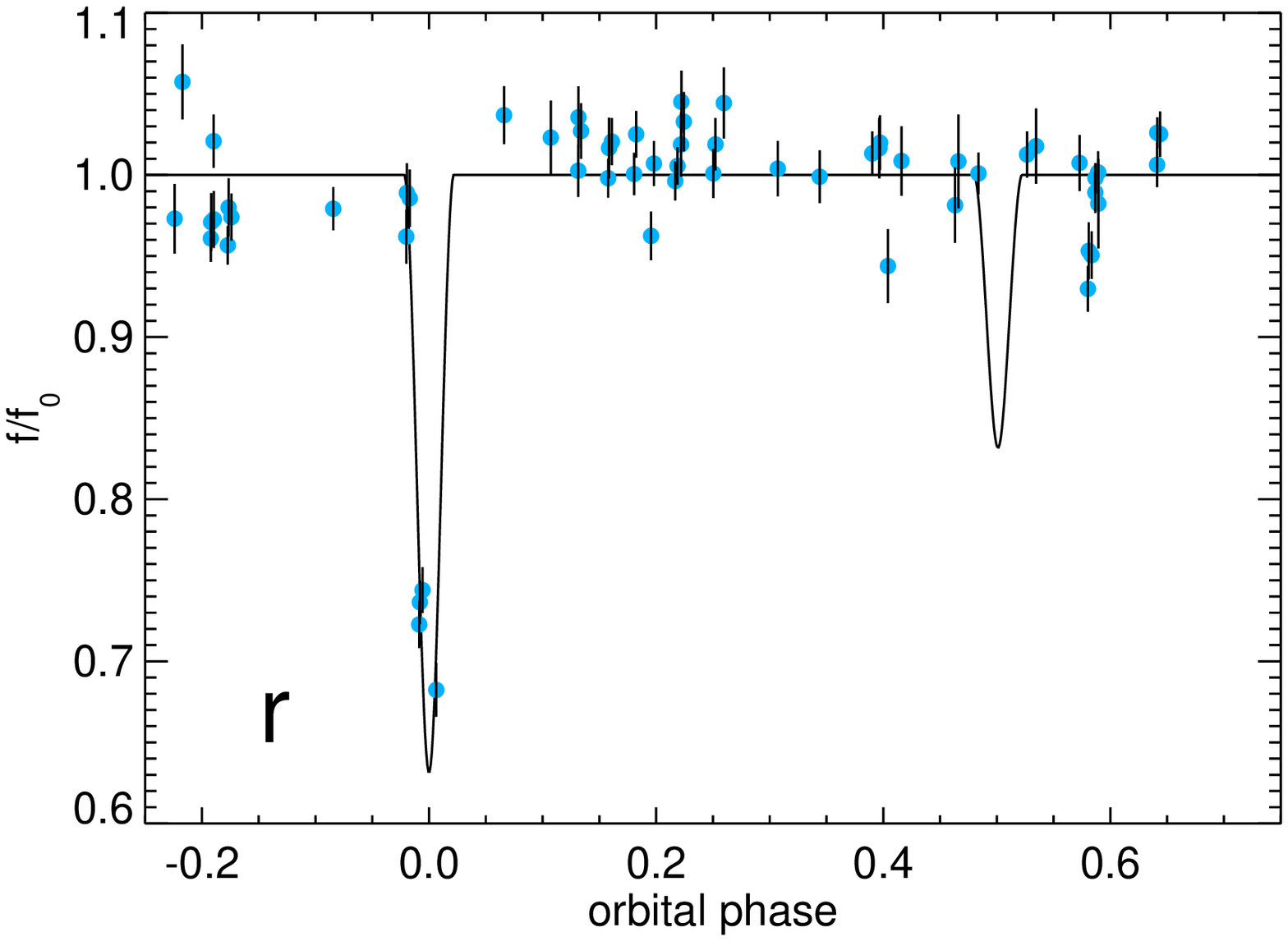}
  \includegraphics[width=8cm]{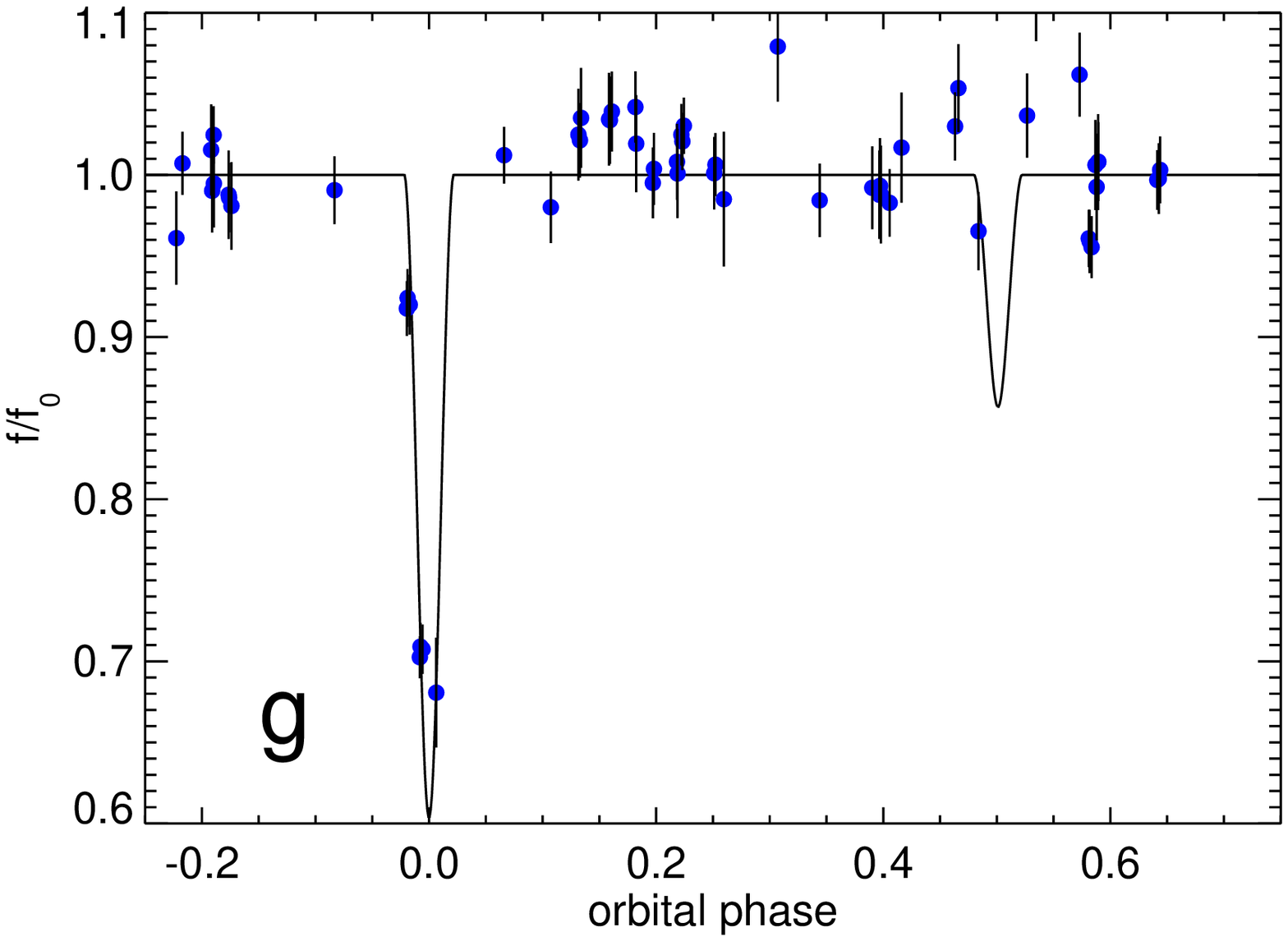}
  \includegraphics[width=8cm]{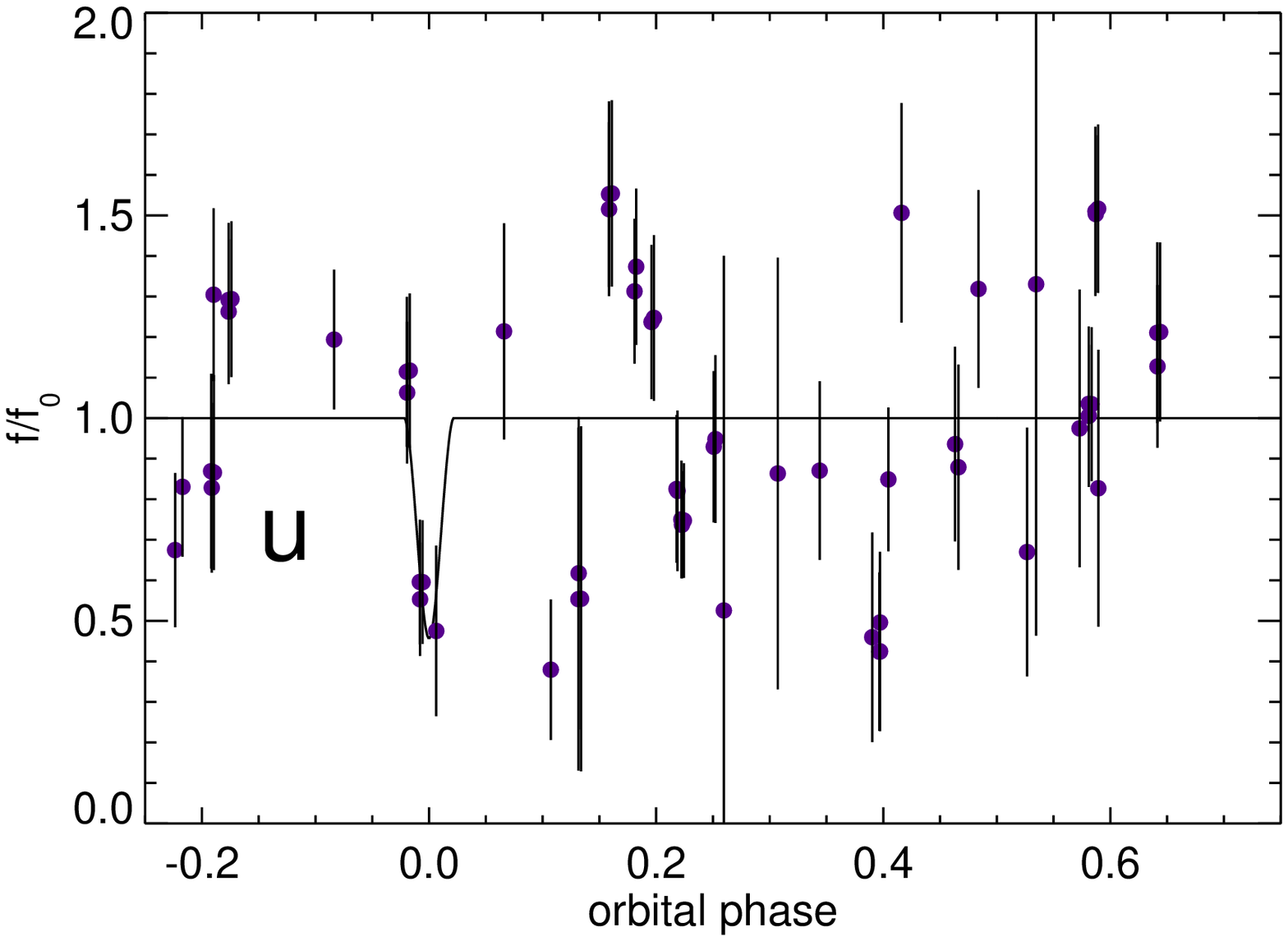}
  \caption{Lightcurve of \name\ folded at the best--fit period of
    2.6390157 days, with J, H and Ks binned every 30 points for clarity.  }
%    The lightcurves are ordered from top to bottom $K_{\rm s}, H, J, z, i, r, g$.}
  \label{fig-lc}
\end{figure*}

Figure~\ref{fig-lc} displays the ensemble lightcurve, folded at the
best--fit period of 2.6390157 days.  The lightcurves are ordered from
top to bottom and left to right $K_{\rm s}, H, J, z, i, r, g, u$.  The
$J$, $H$, and $K_{\rm s}$ data are binned every 30 points.  We note
that SDSS has no data during the secondary eclipse, resulting in
poorly constrained relative optical colours for each star.  This
system is one of the faintest known eclipsing low--mass systems ($r =
18.3$), meaning substantial telescope time is required to measure the
radial velocity curve to high precision.

%We note for completeness that the times of all 2MASS and SDSS
%observations are corrected to the solar system barycenter.

\subsection{Spectroscopy}
\subsubsection{Apache Point Observatory}
\label{subsec-obs:apo}
To confirm our M--dwarf classification, we obtained spectroscopic
observations of this system using the ARC 3.5-m telescope with the
Dual-Imaging Spectrograph (DIS-III) at Apache Point Observatory (APO)
on the nights of 2005 November 22,
%(1 exposure)
2005 November 28,
%(12 exposures), 
and 2005 December 04 UT.
%(15 exposures)
We used the HIGH resolution gratings (0.84\AA/pixel in the red;
0.62\AA/pixel in the blue) and a $1.\arcsec 5$ slit centered at
6800\AA\ (red) and 4600\AA\ (blue).  The chips were binned 2$\times$1
to increase the signal-to-noise and windowed from their original size
of 2048k$\times$2048k to reduce the readout time between
exposures. The approximate wavelength coverage is $\sim 1000$\AA\ in
both the red and blue wavelength regions.
%
%We were unable to obtain useful spectra in the blue region of the
%spectrum due in part to the small total flux in this region and poor
%observing conditions on all three of the nights.

The spectroscopic reductions were performed using standard
$IRAF$\footnote{$IRAF$ is written and supported by the $IRAF$
programming group at the National Optical Astronomy Observatories
(NOAO) in Tucson, AZ.  NOAO is operated by the Association of
Universities for Research in Astronomy (AURA), Inc. under cooperative
agreement with the National Science Foundation.
http://iraf.noao.edu/} reduction procedures.  The calibration images
(bias, flat, arc, flux) used to correct each of the individual spectra
were applied only to images taken on the same night.  The bias, and
flat images were observed at the beginning or end of each night.  A
He--Ne--Ar arc lamp spectrum was taken after each exposure on the
target star.
%
%Because the spectrograph is mounted at the Nasmyth focus and the
%flexure is negligible, this is not entirely necessary but was
%performed regardless.
%

%not sure this is important - AAW

%The average of $\sim 10-15$ bias frames was subtracted from each of
%the individual target spectra.  Each image was then normalized by the
%average of $\sim 5$ bright quartz lamp flat fields.  One to two flux
%and M--dwarf radial velocity spectra were also observed each evening.

%
%It should be noted that due to the non-photometric observing
%conditions, the flux calibration represents a relative flux --- this
%is perfectly acceptable because we require only relative flux values
%between the various points in the stellar spectrum for our purposes.
%

\begin{figure}
  \includegraphics[width=8cm]{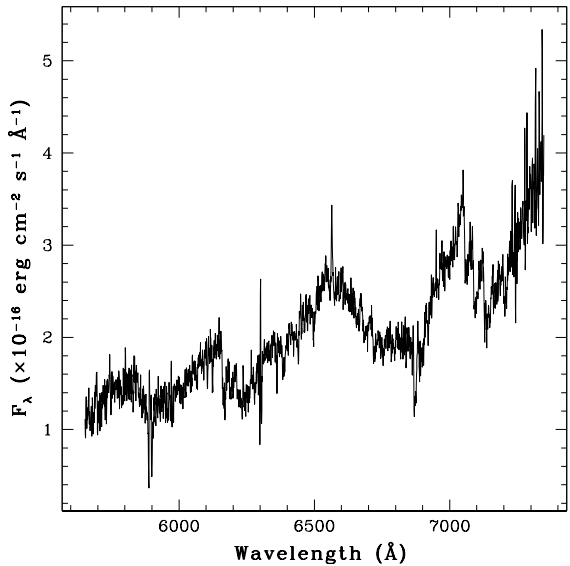}
  \caption{Spectrum of \name\ obtained on the ARC 3.5-m telescope with
    the DIS-III spectrograph (0.84\AA/pixel) on 2005 November 22 UT.
    Note the H$\alpha$ emission near 6563 \AA, indicating chromospheric
    activity.}
  \label{DISspec}
\end{figure}

A representative spectrum of \name\ is displayed in Figure
\ref{DISspec}.  These initial spectra clearly demonstrate the features
of an M--dwarf system.  The presence of H$\alpha$ in emission at 6563
\AA\ indicates magnetic activity, although no obvious line splitting
is detected.  The broad molecular bands of TiO ($\sim 7050$ \AA) are
also readily apparent.

%With these data alone, it is uncertain if the H$\alpha$ emission is
%magnetically induced in one star or both, or a product of interaction.

\subsubsection{Magellan}
\label{subsec-obs:mag}

%We obtained two 600 second exposures with the Low Dispersion Survey
%Spectrograph 3 (LDSS3) on the Magellan II telescope on the night of
%December 29, 2005.  The VPH Blue grating (0.682 \AA/pixel; R = 1900)
%and a $2\arcsec$ slit were used.  The spectrum was processed like...
%subtracted a combined bias, applied overscan correction, flat-fielded
%with (quartz lamp?) flatfield, cleaned the image of cosmic rays,
%extracted a 20 pixel wide region centered on the target star, used a
%(quartz) lamp spectrum to derive the dispersion solution.  The final
%spectrum has a wavelength range of $4500\AA - 7000\AA$

Two 600 second exposures of \name\ were obtained with the Low
Dispersion Survey Spectrograph 3 (LDSS3) on the Magellan II/Clay
telescope on the night of December 29, 2005.  The VPH Blue grating
(0.682 \AA/pixel; R = 1900) and a $2\arcsec$ slit were used.  The
spectra were reduced and calibrated employing standard techniques,
which include subtracting a combined bias, subtracting the overscan
correction, flat-fielding with quartz lamp observations, cleaning the
image of cosmic rays, and extracting a region centered on the target
star.  The dispersion solution was derived from an observation of a
He--Ne--Ar arc lamp.
%
%The final spectrum has a wavelength range of $XXXX\AA - XXXX\AA$.
%
Flux calibrations were derived from observations of spectrophotometric
standard stars from \cite{Oke-83}.  The spectra were corrected for the
continuum atmospheric extinction using mean extinction curves.
Telluric lines were removed using a procedure similar to that of
\cite{Wade-88} and \cite{Bessell-99}.  We use the Magellan spectra to
estimate the spectral types of the components in
Section~\ref{sec-templatematch}.

%The resulting spectrum is a higher S/N than the ARC observations, and
%definitively show TiO bands at XXXX \AA.  

\subsubsection{Keck}
\label{subsec-obs:keck}
We were unable to resolve line splitting in either the APO or Magellan
spectra, and therefore made use of the HIRES spectrograph at Keck
during observing runs on 2006 October 13, 2006 December 12 and 2007
January 6.  Over the 3 nights we obtained five spectra at
R$\sim$50,000 with exposure times ranging from 30-40 minutes each.

\begin{figure}
  \includegraphics[width=8cm]{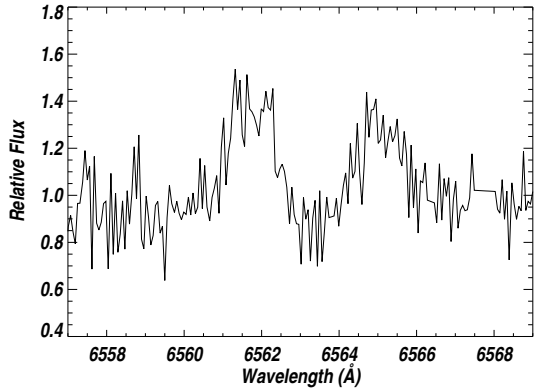}
  \caption{Spectrum of \name\ obtained with HIRES (0.06 \AA/pixel) on
    the Keck 10-m telescope on Oct 13, 2006.  Note the line splitting
    around H$\alpha$ at 6563 \AA, yielding emission peaks centered near
    6561.5 \AA\ and 6564.9 \AA.}
  \label{fig-keck}
\end{figure}

\begin{figure}
  \includegraphics[width=8cm]{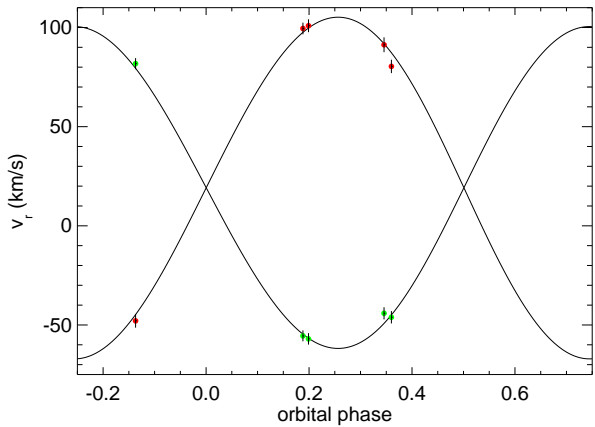}
  \caption{Radial velocity curve derived from the Keck data.  The system
    velocity is $19.1 \pm 1.3$ km s$^{-1}$.  Velocities are relative to
    the solar system barycenter.}
  \label{fig-rv}
\end{figure}

The data were reduced using standard IDL routines that included order
extraction, sky subtraction, and cosmic ray removal.  The resulting
S/N per pixel (0.06 \AA/pixel) ranged from 6-9.  The detection of
H$\alpha$ emission lines in both stars due to magnetic activity
allowed measurement of their radial velocities (we were unable to use
cross correlation techniques on the lower resolution APO and Magellan
spectra).  An illustrative example of the Keck data is shown in Figure
\ref{fig-keck}.  The H$\alpha$ emission lines were fit to a double
Gaussian profile using Levenberg--Marquardt least--squares
minimization. Epochs and radial velocities for each of the 5
observations can be found in Table~\ref{tab-rvs}.  The radial velocity
curve derived from these data is shown in Figure~\ref{fig-rv}.

\section{Analysis}
\subsection{Modeling The System}\label{sec-model}
We corrected the times of all observations, as well as measured radial
velocities, to the solar system barycenter (TDB).  We analyzed the lightcurve 
of \name\ using the code of \cite{2002ApJ...580L.171M}.  All of the orbital
elements of the binary were allowed to vary, and we allowed the masses
and radial velocity of the center of mass of the system to vary
simultaneously while fitting the parameters describing the lightcurve.
%Since \name\ is a double--lined spectroscopic binary, we were able to
%derive the masses, radii and fluxes of both stars independent of any
%stellar atmosphere model.  
As both stars are well within their Roche radii and the centripetal
acceleration at the equator is three orders of magnitude smaller than
the surface gravity, we treated each star as a sphere (and hence
ignore gravity darkening). Also, the stars are sufficiently separated
that we can ignore reflected light.  The one uncertainty in our models
is the limb-darkening, which should be modest in the infrared
\citep{1995A&AS..114..247C}.  We find that the assumed limb-darkening
affects our results very little, so we fix the linear limb-darkening
coefficients for each star, $u_{1,2}$, at the values computed by
\citet{2000A&A...363.1081C,2004A&A...428.1001C} for model atmospheres
with $T_{eff}=3800$ and $3600$ for the primary and secondary,
respectively.  We assume $log(g)=4.5$, $[M/H]=0$, and a microturbulent
velocity of $2$ km s$^{-1}$ for both, based on the {\it PHOENIX}
atmosphere models of \cite{1999ApJ...512..377H}. The limb-darkening
parameters are given in Table \ref{tab-pars1}.

Our model contains 25 free parameters, starting with the five orbital
parameters\footnote{The longitude of the ascending node is
unconstrained since the reference plane is the sky plane and nothing
in our data constrains the position angle of the binary on the sky.}:
the eccentricity and longitude of pericenter combine to two
parameters, $e \cos{\omega}$ and $e \sin{\omega}$; the inclination,
$i$; the time of primary eclipse $T_0$ (which can be translated into a
time of periapse); and the period, $P$ (which can be translated into a
semi--major axis from the total mass of the system).  Four parameters
describe the bulk stellar properties: $R_1$, $R_2$, $M_1$, and $M_2$.
The radial velocity of the center of mass of the system is $\gamma$
(in km s$^{-1}$).  The fluxes are described by 15 parameters (we hold
the $u$-band flux of the second star fixed at zero as the best--fit
value is negative).  For the model fitting, we transformed to the set
of parameters suggested by \citet{2006MNRAS.367.1521T} which have
weaker correlations between the transformed parameters.  We found the
initial best--fit model using Levenberg--Marquardt least--squares
non--linear optimization giving a best-fit model with $\chi^2=11304.4$
for 9168 degrees of freedom.  We found that the scatter of the data
outside of eclipse had a gaussian shape, but with a larger scatter
than the errors would warrant.
%, about 10\% larger in the infrared, and 50-200\% larger in the SDSS bands.  
It is possible that this discrepancy is due to variability in the
stellar fluxes as the data were gathered over several years, or that
the error bars are simply underestimated, so we added a systematic
error in quadrature ($0.04, 0.03, 0.03, 0.01, 0.02, 0.02, 0.01, 0.26$
magnitudes in the $K_{\rm s}, H, J, z, i, r, g, u$--bands,
respectively) such that the reduced $\chi^2$ of the data outside of
eclipse in each band is equal to unity, and then refit the entire data
set.  The resulting $\chi^2$ of 9253.2 has a formal probability
$P(\chi^2>9253.2)=$26\% for 9168 degrees of freedom.  The best fit
parameters for the brightness of the system 
are given in Table~\ref{tab-pars1} and
the orbital and physical parameters in Table~\ref{tab-pars2}.

As the number of free parameters is large and the parameters can be
strongly correlated, we ran simulated data sets, adding gaussian noise
to the best-fit lightcurve and radial velocity model values.  We ran
$10^5$ simulations, and for each simulation re-ran the fitting routine
to derive the best-fit parameters from the simulated data. We used
the $10^5$ sets of derived parameters simulations to compute the 
1-$\sigma$ errors on the best-fit model parameters, sorting each parameter
and choosing the 68.3\% confidence intervals, which are given in 
Tables~\ref{tab-pars1} and \ref{tab-pars2}.  We computed a 90\% confidence 
upper limits on the $u$ band flux of Star 2 by increasing its value from zero,
while minimizing over the other system parameters, until the change in
chi-square was $\Delta \chi^2=2.71$.  Our analysis yields asymmetric
error bars for all system parameters.  We quote a single error bar
when the negative and positive uncertainties are the same to within
$10\%$.

The best--fit lightcurve is shown in Figure~\ref{fig-lc}.  The total
flux of the two stars is very well constrained due to the large number
of data points of high photometric quality.  However, the individual
fluxes are more poorly constrained due to the small number of points
taken during the eclipses, so that uncertainties in the radius,
limb-darkening, and inclination lead to uncertainties in what fraction
of each total stellar flux is obscured during primary and secondary
eclipse.  Thus the derived fluxes of each star are strongly
anti-correlated, which is why the errors on the individual fluxes are
much larger than the error on the total flux (Table~\ref{tab-pars1}).
The fit indicates that, in the infrared passbands, the eclipses are of
very similar depths (0.35 vs 0.31 magnitudes in the $Ks$--band), which
is the reason that the original Supersmoother fit converged on an
alias of half the period.

\begin{figure*} 
  \includegraphics[width=8cm]{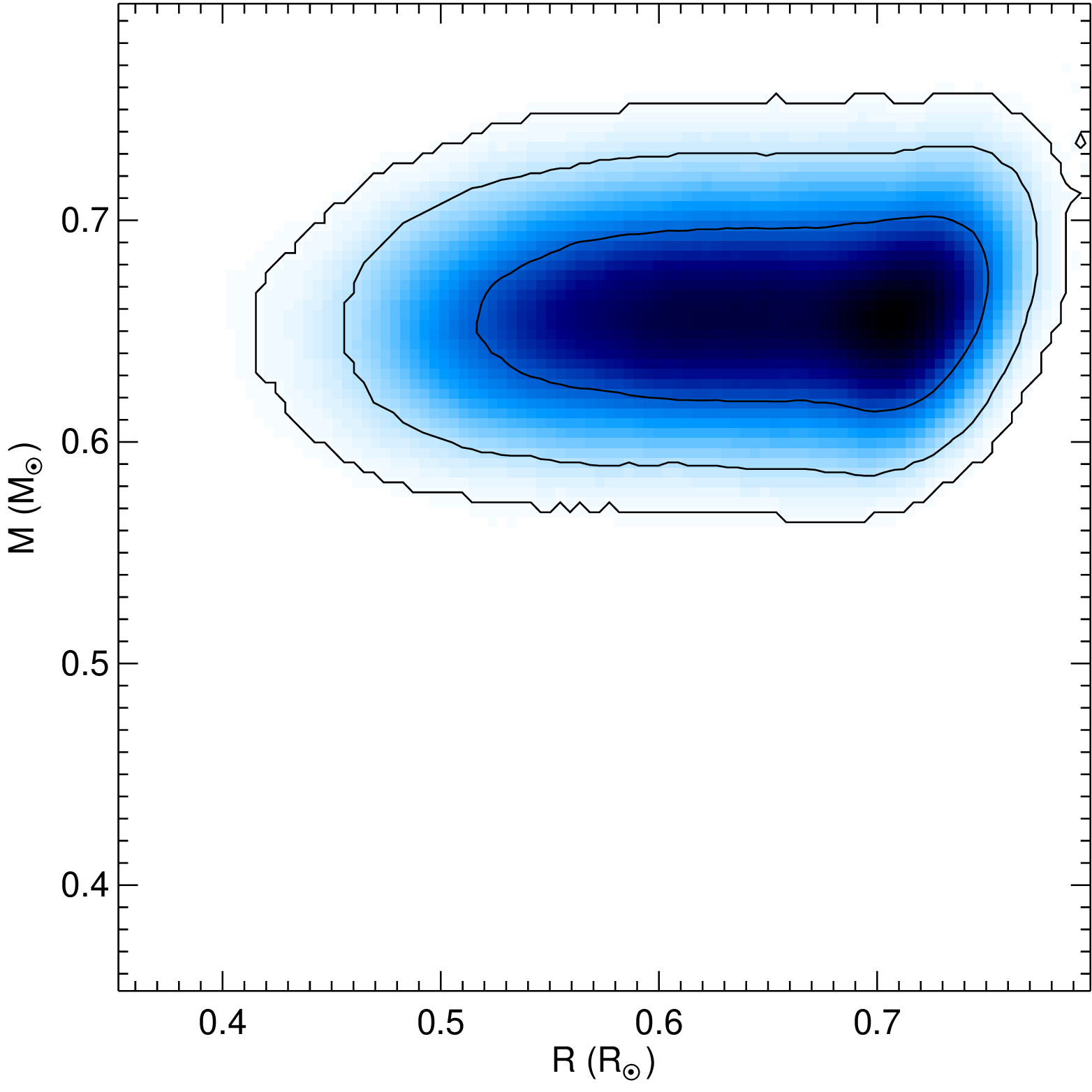}
  \includegraphics[width=8cm]{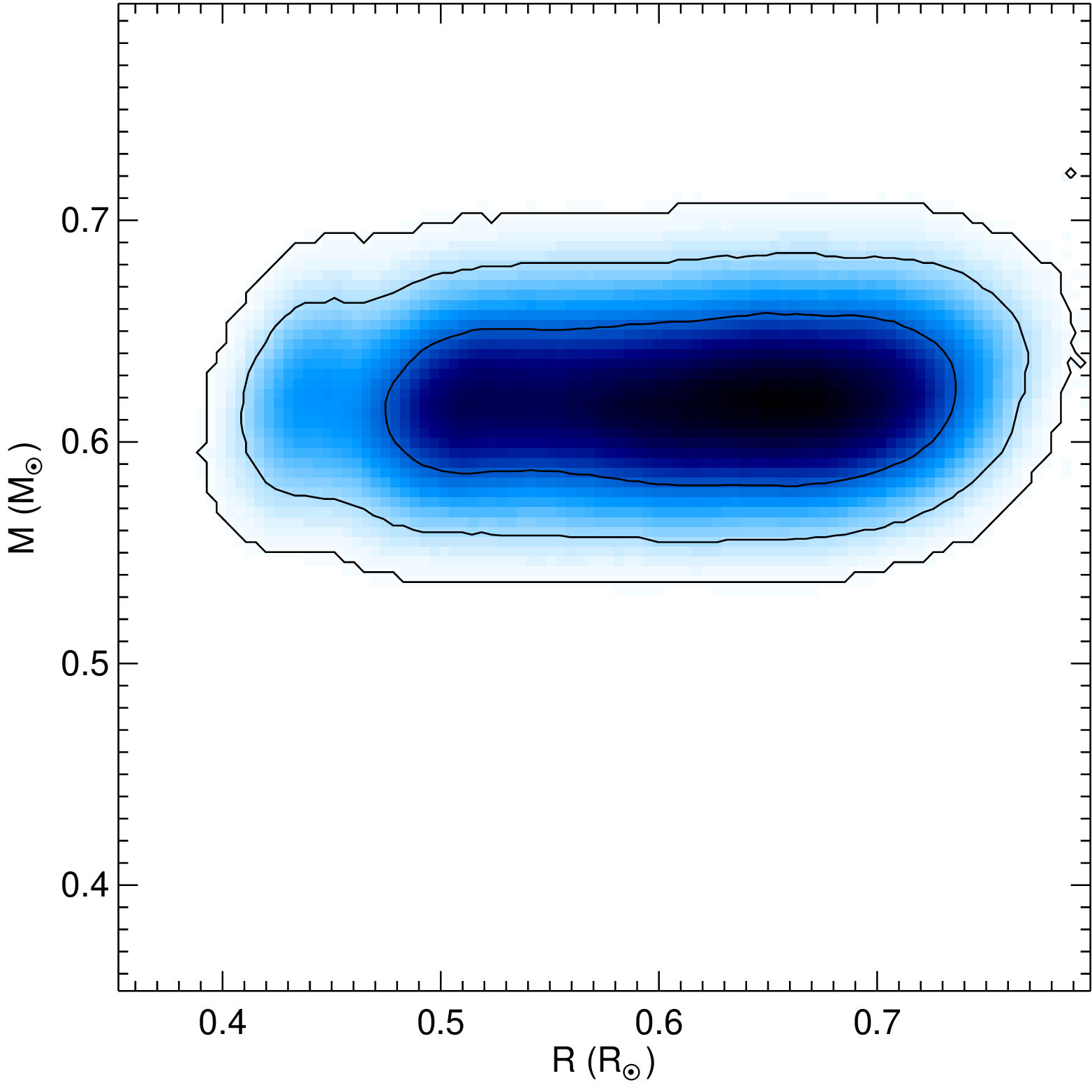}
  \caption{Mass--radius parameter space allowed by our data on \name.
    The left figure is for the heavier (primary) object, while the right
    figure is for the secondary.  The shaded area correspond to the
    probability distribution derived from the simulated data sets,
    while the contours represent the 1-,2-, and 3-sigma
    confidence regions.}
  \label{fig-mr} 
\end{figure*}

The allowed mass--radius parameter space for each star is shown in
Figure~\ref{fig-mr}.  Each panel shows the probability distribution of
the mass and radius of each star derived from the simulated data sets
in units of solar radius and mass.  The contours are 1-,2-, and
3-sigma confidence regions (68.3\%, 95.4\%, and 99.73\% of the $10^5$
parameter sets).  We compare the derived values to the masses and
radii found for other low--mass stars in Section~\ref{sec-ana}.

From our results, we can derive several auxiliary parameters for the
stars which have different error bars than if they are computed from
the parameters in Table~\ref{tab-pars2} due to covariance between
model parameters.  We also list in Table~\ref{tab-pars2} (below the
horizontal line) : the mass-radius ratio for each star,
$M_{1,2}/R_{1,2}$; the ratio of the radii of the two stars, $R_1/R_2$;
the sum of the stellar radii, $R_1+R_2$; the ratios of the semi-major
axis to the stellar radii, $a/R_{1,2}$; the velocity semi-amplitudes,
$K_{1,2}$ and the total amplitude $K$; the surface gravities,
$log(g_{1,2})$; the stellar densities $\rho_{1,2}$; and the total mass
of the system, $M_1+M_2$, which can be used to derive a semi-major
axis of the system, $a$.  At mid-eclipse, the projected separation of
the stellar centers on the sky is $b$.  The fractional error on
$(R_1+R_2)/a$ is much smaller ($\sim 2$\%) than on $R_1$ or $R_2$
individually ($> 10$\%) due to strong correlations between $R_1/a$,
$R_2/a$ and $i$, as discussed by \citet{2006MNRAS.367.1521T}.  Since
our derived fluxes of the stars cover the peak in their spectral
energy distributions, we have derived the bolometric flux ratio of the
stars by smoothly interpolating between the fluxes in the different
bands, and then taking the ratio of the two stars.  Given the relative
sizes and fluxes of the stars, we derive the ratio of the effective
temperatures, $T_2/T_1$ (we cannot derive the absolute fluxes or
effective temperatures from the lightcurve and RV data as we do not
know the distance to this system to high precision).

\subsection{Spectral Types and Temperature Estimates}
While the masses, radii and fluxes for each star are well-determined
by our modeling procedure, there are other quantities that can be
measured from our observations.  The spectral types of each component,
total space velocity of the system and effective temperature estimates
can be determined from our assembled dataset.
	
\subsubsection{Binary Spectral Template Matching}
\label{sec-templatematch}
\begin{figure}
  \includegraphics[width=8cm]{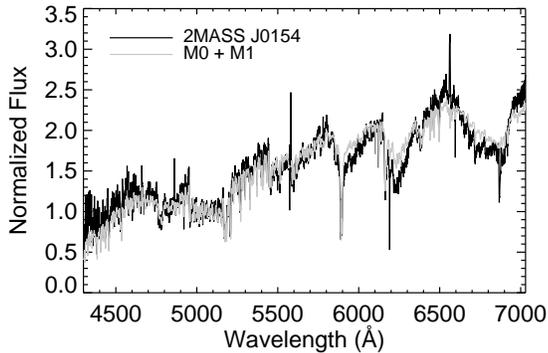}
  \caption{Magellan spectra (0.682 \AA/pixel) of \name\ along with the
    best fit binary spectral template.  The best fit binary template is
    composed of an M0 + M1.}
  \label{fig-magellan}
\end{figure}

In order to estimate the spectral types of this composite system, we
constructed a grid of binary spectral templates.  The spectra employed
in synthesizing the binary templates were drawn from the low--mass
stellar templates of \cite{2007AJ....133..531B} and the K5 and K7
templates used in the HAMMER spectral analysis software package
\citep{2007arXiv0707.4473C}.  Each template spectrum was scaled by its
bolometric luminosity \citep{rei05} and coadded with all other
templates.  Relative velocity shifts were introduced for each spectral
type pair, ranging from -200 km s$^{-1}$ to 200 km s$^{-1}$ in steps
of 20 km s$^{-1}$.  The final binary spectra grid consisted of 1638
templates.  The templates were normalized to the Magellan observations
and residuals were computed from 4500 to 7000 \AA.
%\footnote{Residuals
%were computed as:
%\begin{equation}
%\chi^2 = \Sigma(flux_{\lambda Obs.} - flux_{\lambda Template})^2
%\end{equation}
%}.  
The best fit binary pair for the Magellan spectra is an M0 primary and
M1 secondary, with an uncertainty of $\pm 1$ subtype for each
component.  We did not use the velocity information as a constraint,
but rather included the relative shifts for completeness.  We adopt
these subtypes for each component.  The best fit composite spectrum is
shown in Figure~\ref{fig-magellan}, along with the Magellan data.

\subsubsection{Spectral Types From Optical--IR Colours}
Using the colour--spectral type relationships for $r-i$, $i-z$, and
$i-J$ derived by \cite{West05a} and \cite{2007AJ....133..531B}, we
estimate a spectral type of M0 ($\pm$ 1 subtype) for the primary and
M3 for the secondary. The intrinsic spread in colour at a given
spectral type and our measured errors permit a range of spectral types
between M0 and M4 for the secondary.
Because of this large uncertainty, we adopt the secondary subtype
derived from the spectral template analysis, and note that these
colour--based results are consistent with the spectroscopic results
derived in Section~\ref{sec-templatematch}.
%
%These spectral types are consistent with those derived from the
%spectra above.  
%
We estimate a distance of $623 \pm 50$ pc from the $i$--band
photometric parallax \citep{West05a} of the primary star.  At a
Galactic latitude of $b$ = $-58 \deg$, this binary has vertical
distance below the Disk of $\sim 530$ pc, consistent with being a
member of the Galactic thin disk.

% COLOURS
%[$r-i; i-z; i-J$] colours of the stars we
%estimate spectral types for the primary star of 
%% [0.89; 0.48; 1.60] 
%[M1; M1; --] and for the secondary of
%% [1.08; 0.66; 2.24]
%[M2;M2--M3;M3--M4].  We adopt a spectral type of M1 for the primary,
%and of M3 for the secondary.  These classifications yield distances of
%741 and 711 pc for the primary and secondary.

% BRIGHTNESS
% J=band distance : 9.34 and 9.26 magnitudes = 5 log r - 5 = 741 and 711 pc for the primary and secondary.
% 
% i-band for i-z=0.48; M_i = 7.18 + 3.14(i-z) +/- 0.08
% absolute mag = 8.6872; apparent mag = 17.66; dist mod = 8.9728 = 623 pc
% 8.9728 + sqrt(0.08**2 + 0.15**2) = 9.1428 = 673pc
% 8.9728 - sqrt(0.08**2 + 0.15**2) = 8.8028 = 576pc

The thin disk membership of \name\ is strengthened by examining the
system's kinematics.  The tangential velocity implied by the observed
proper motion ($11.2 \pm 3.2$ mas year$^{-1}$) and distance to the
system is $33.1 \pm 12.1$ km s$^{-1}$.  Added in quadrature with the
system radial velocity from the binary fit, we find a space velocity
of $38.2 \pm 12.2$ km s$^{-1}$, again consistent with the thin disk
\citep{2007arXiv0708.0044B}.

% 11.2 milliarcsec = 5.42991322843e-08 radians
% 1 year           = 31557600.0 s
% 11.2 mas / yr    = 1.72063567205e-15 rad/s
% 3.2 mas / yr     = 4.91610192014e-16 rad/s

% 623pc = 1.921955e16 km
% 50pc = 1.5425e+15

% velocity = 33.0698433307 km/s +/- 12.1026071901

\subsubsection{Metallicity and Temperature Estimates}
\label{subsec-spec:metal}
We measured the composite CaH2, CaH3, and TiO5 molecular indices in
our APO spectra of this system, and find CaH2 = $0.64 \pm 0.06$, CaH3
= $0.83 \pm 0.03$, and TiO5 = $0.64 \pm 0.04$. Using these values
along with the empirical formula from Figure~2 of \cite{WoolfWall06}
yields an effective temperature estimate for the primary of $T_{\rm
eff} = 3730 \pm 100K$, consistent with the M0 spectral type determined
from the full spectra.  Further, when these measurements are compared
with the samples of \cite{2003ApJ...585L..69L}, \cite{WoolfWall06},
\cite{2006ApJ...645.1485B}, and \cite{West07a}, they suggest that the
composite system is of solar or slightly super--solar metallicity,
again consistent with thin disk membership\footnote{It should be noted
that there exist dMs in \cite{WoolfWall06} with similar CaH and TiO
indices to our targets, but with [Fe/H] values below $-0.3$}.

% screw this; we need 2 colour transformations!
% SDSS -> Cousins
% 2MASS -> CIT
%
%The 2MASS magnitudes may be related to the Caltech (CIT) photometric
%system magnitudes through \cite{Carpenter-01}\footnote{Updated
%transformations are found at
%\url{http://www.astro.caltech.edu/~jmc/2mass/v3/transformations/}}.
%For the reason of checking mass ($\msun$) -- colour ($V - K$) relations
%from \cite{Delfosse-00}, who use Johnson--Cousins--CIT magnitudes.

\section{Discussion}

\begin{figure}
  \includegraphics[width=8cm]{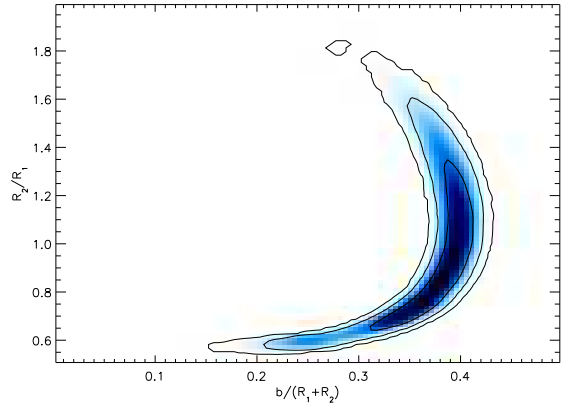}
  \caption{Confidence limits on $b/(R_1+R_2)$ and $R_2/R_1$ from the
synthetic data sets.  The shaded region shows the density of points
from the best-fit parameters to $10^5$ simulated data sets, while
the contours are 68.3\%, 95.4\%, and 99.73\% confidence regions.}
  \label{fig-banana}
\end{figure}

Previous studies \citep[e.g.][and references
therein]{2006Ap&SS.304...89R}, have demonstrated that current models
underpredict the radii of low--mass stars at a given mass.  The radii
derived for \name's components (R$_1 = 0.64 \pm 0.08 \rsun$ at M$_1 =
0.66 \pm 0.03 \msun$, R$_2 = 0.61 \pm 0.09 \rsun$ at M$_2 = 0.62 \pm
0.03 \msun$) lie in between different model predictions, and the
errors are currently large enough to not yet provide discrimination
between the models.
%
%Nearly the entire source of the uncertainties on the stellar radii is
%due to the large errors on the photometry due the faintness of this
%system.
%
The source of the uncertainties on the stellar radii is almost
entirely due to the large errors on the photometry given the faintness
of this system.
A severe banana-shaped degeneracy between the impact-parameter (or
inclination) and the ratio of the stellar radii occurs due to the
large photometric errors (Figure \ref{fig-banana}); this is why in
Table \ref{tab-pars2} the fractional error on $R_1+R_2$ ($\sim$ 3\%)
is much smaller than the fractional error on $R_2/R_1$ ($\sim$ 30\%).
Within the 68.3\% confidence limit, the deviation of the K-band
lightcurve from the best-fit lightcurve is only 0.6\%, which implies
that milli-magnitude precision would be required to derive the radius
ratio to high accuracy.  Below, we discuss the implications of our
system with regards to current theoretical and empirical mass-radius
relations.

\subsection{The Empirical Mass--Radius Relationship}\label{sec-ana}
\begin{figure}
  \includegraphics[width=8cm]{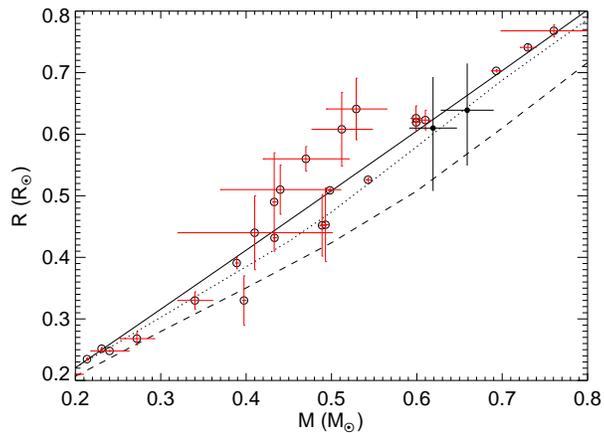}
  \caption{Distribution of mass vs. radius for known low--mass
    double--lined eclipsing binary systems (open circles;
    L{\'o}pez-Morales 2007; L{\'o}pez-Morales et~al. 2006;
    L{\'o}pez-Morales \& Shaw 2007; Maxted et~al. 2004).  The solid
    line shows an empirical relationship derived by Bayless \& Orosz
    (2006); the dotted line represents the Z=0.02, Y=0.275, T=1 Gyr
    models of Baraffe et~al. (1998); the dashed line represents the
    Z=0.02, Y=0.28, T=1 Gyr models of Siess et~al. (2000).  The solid
    dots correspond to the components of \name.}

    %double--lined eclipsing binary systems \citep[open
    %circles;][]{2007ApJ...660..732L,2006astro.ph.10225L,2007ASPC..362...26L,2004MNRAS.355.1143M}.
    %The solid line shows an empirical relationship derived by
    %\cite{2006ApJ...651.1155B}; the dotted line represents the Z=0.02,
    %Y=0.275, T=1 Gyr models of \cite{1998A&A...337..403B}; the dashed
    %line represents the Z=0.02, Y=0.28, T=1 Gyr models of
    %\cite{2000A&A...358..593S}.  The solid dots correspond to the
    %components of \name.}
  \label{fig-mrdist} 
\end{figure}

In Figure ~\ref{fig-mrdist}, the masses and radii of known low--mass
eclipsing binary systems
\citep{2007ApJ...660..732L,2006astro.ph.10225L,2007ASPC..362...26L}
are plotted along with current empirical \citep{2006ApJ...651.1155B}
and theoretical models
\citep{1998A&A...337..403B,2000A&A...358..593S}.
\cite{2006ApJ...651.1155B} have derived an empirical mass--radius
relationship for K and M dwarfs from known binaries that stretches up
to nearly $0.8 \msun$.  We test this empirical mass-radius relation by
comparing their expected radii, given our mass measurements, to our
measured radii.  Their analysis predicts $R_1 = 0.67 \pm 0.03 \rsun$
and $R_2 = 0.63 \pm 0.03 \rsun$, while we measure $R_1 = 0.64 \pm 0.08
\rsun$ and $R_2 = 0.61 \pm 0.09 \rsun$.  Our objects are consistent
with the ensemble of eclipsing binary stars used to derive their
relationship, and fill in the gap at the high--mass, early--dM end of
the relationship.  The dearth of data in Figure~\ref{fig-mrdist} and
the recent discovery of many of these systems reflects that this is an
emerging field, only recently enabled by large--scale photometric
surveys.

Theoretical models predict mass-radius relations which are a strong
functions of both metallicity and age \citep{1996ApJ...461L..51B}.
The predictions of the \cite{1998A&A...337..403B} evolutionary models
for objects of solar metallicity and ages of $10^8-10^9$ years are
consistent with \name, yet disagree with other systems of similar
mass.  The models tend to systematically underpredict the radii at a
given mass, as shown in Figure~\ref{fig-mrdist}.  While the
uncertainties on the masses and radii of \name's components render
them an equally good fit to both the \cite{1998A&A...337..403B}
theoretical models and \cite{2006ApJ...651.1155B} empirical fit,
additional photometric and spectroscopic data should yield more
precise measurements of these attributes, and better discrimination
between models.

For comparison, the models of \cite{2000A&A...358..593S} are also
shown in Figure~\ref{fig-mrdist}.  The large differences between
models with similar inputs for metallicity and age highlight the
uncertainty that presently exists in this field.  Hopefully this
situation will be remedied by more high-precision measurements of
fundamental stellar parameters in binary systems, along with updated
models.

%Some puzzling discrepancies appear when comparing \name\ with other
%similar stars.  The derived masses of \name\ are among the largest
%measured for known M--dwarf systems \citep{rei05}.
%\cite{2002ApJ...567.1140T} find masses of $0.5992\pm 0.0047 M_\odot$
%and radii of $0.6191 \pm 0.0057 R_\odot$ for both components of YY
%Gem, which they classify as an M1.0 Ve star.  The primary star of
%\name\ has a similar spectral type as YY Gem, but its mass and radius
%are larger than those of YY Gem at the $2-\sigma$ level, while the
%secondary star of \name\ has a later spectral type than YY Gem, but
%its mass and radius are consistent with that of YY Gem at the
%$1-\sigma$ level.  Additional photometry and radial velocity
%measurements of \name\ will be necessary to determine if these
%differences are statistically significant.

\subsection{Activity}
An interesting aspect of this system is the observed H$\alpha$
emission in both of the components.  \cite{2004AJ....128..426W} find
that less than 5\% of isolated M0 and M1 stars show activity
(H$\alpha$ equivalent width of at least 1\AA).  The activity in
early-type M dwarfs is also short lived \citep[$<$ 1
Gyr;][]{2000vlms.conf..109H,West07a}.  It would be very unlikely to
randomly draw an active M0 and M1 star from the field population,
suggesting that some aspect of the interactions between the components
is inducing the observed magnetic activity.  The improbability of the
stars being independently active is furthered when we consider the
distance that this pair is from the Galactic plane.  It is likely that
M0 and M1 dwarfs that are several hundred pc from the Plane have been
dynamically heated for many Gyrs and have ceased being active
\citep{West07a}.

A search through the Palomar/MSU Nearby Star Catalog of
\cite{2002AJ....123.3356G} shows that a large fraction (20/22) of
double--lined spectroscopic M--dwarf binaries have magnetically active
components (H$\alpha$ equivalent width of at least 1\AA).  Other
empirical studies \citep[e.g.][]{2007ApJ...660..732L} have suggested
that the magnetic activity and metallicity of a star may affect its
radius, drawing an explicit connection between the enhanced activity
and large radii found in M--dwarfs in binary systems.
\cite{2007A&A...472L..17C} suggest that enhanced magnetic activity may
lead to inefficient thermal transport in stellar interiors.  This
results in objects with larger radii and smaller effective
temperatures than stars where magnetic effects are negligible.  Rapid
stellar rotation may similarly affect the interior convection.  In
addition, enhanced surface spot coverage ($30-50\%$), due to strong
magnetic activity, may impact the stellar radius.  Either of these
effects may be responsible for the empirical mass--radius
relationships found in low--mass binary stars; additional data are
required to constrain these theories.

%\cite{2007A&A...472L..17C} suggest that the enhanced magnetic activity
%associated with stars in close binary systems may lead to inefficient
%thermal transport in the stellar interiors.  This results in objects
%with larger radii and smaller effective temperatures than stars where
%magnetic effects are negligible.  Rapid stellar rotation may similarly
%affect the interior convection.  \cite{2007A&A...472L..17C} also
%indicate that enhanced surface spot coverage ($30-50\%$), due to
%strong magnetic activity, is also sufficient to explain the empirical
%trends.

\section{Conclusions}
\label{sec-conclude}
We report the discovery and characterization of the double--lined
eclipsing binary system \namefull.  Photometric and spectroscopic
evidence suggests that both components are M dwarfs, and we adopt
classifications of M0 and M1 as their subtypes.  We resolve splitting
of the H$\alpha$ emission line with spectroscopic observations using
HIRES at Keck, leading to a radial velocity curve and estimates of the
masses and radii of each star.  Simulated data sets created by
adding noise to the best fit provides uncertainties on and covariances 
between the system
parameters.  We emphasize that there exist complicated degeneracies
between parameters in eclipsing systems that can only be fully
explored with such detailed analyses.

Our analysis is consistent with previous studies of double--lined
eclipsing M--dwarf systems.  An empirical study by
\cite{2006ApJ...651.1155B} yields a quadratic mass--radius
relationship spanning the range of $0.2 - 0.8 \msun$ and spectral
types from late K to late M.  This empirical relation accurately
predicts the radii of \name's components.

%We are unable to measure the temperature of the system's components to
%high precision.  However, there are well--known problems in
%establishing the temperatures of M--dwarfs, since M--dwarf spectra are
%poorly approximated by a blackbody curve.  \cite{1992ApJ...392L..31B}
%establish a broadband ``blackbody equivalent temperature'', although
%the uncertainties in this measurement are equivalent to our
%uncertainties on the temperature derived from the CaH2 index
%\citep{WoolfWall06}.  A more useful method would be to determine the
%absolute luminosity of the components, requiring an accurate parallax
%distance.  At a distance of $\sim 600$pc, such observations would
%require 2 mas resolution, easily within reach of the next--generation
%astrometry satellite GAIA \citep{2001A&A...369..339P}.

% (resolving
%orbital motion of the binaries would require resolution of 64
%$\mu$arcseconds, for which these objects are too faint).

We observe H$\alpha$ emission from both components, an unlikely
scenario given their early--M spectral types and their distance from
the Galactic plane.  If magnetic activity is enhanced in M dwarfs in
binary systems, the binary population including M--dwarfs components
may present an additional foreground of stellar flares for next
generation time domain surveys
\citep{2006ApJ...644L..63K,2004ApJ...611..418B}.

\section*{Acknowledgments}
We thank M. Claire, S. Hawley, and M. Solontoi for useful discussions,
and H. Bouy for assistance with the Keck observations.
E. A. acknowledges support from NSF CAREER grant AST-0645416.
G. B. thanks the NSF for grant support through AST-0098468.  This
paper includes data gathered with the 6.5 meter Magellan Telescopes
located at Las Campanas Observatory, Chile.  This publication also
makes use of data products from the Two Micron All Sky Survey, which
is a joint project of the University of Massachusetts and the Infrared
Processing and Analysis Center/California Institute of Technology,
funded by the National Aeronautics and Space Administration and the
National Science Foundation.  This work is based on observations
obtained from the W. M. Keck Observatory, which is operated as a
scientific partnership among the California Institute of Technology,
the University of California, and the National Aeronautics and Space
Administration.

Funding for the Sloan Digital Sky Survey (SDSS) and SDSS-II has been
provided by the Alfred P. Sloan Foundation, the Participating
Institutions, the National Science Foundation, the U.S. Department of
Energy, the National Aeronautics and Space Administration, the
Japanese Monbukagakusho, and the Max Planck Society, and the Higher
Education Funding Council for England. The SDSS Web site is
http://www.sdss.org/.  

The SDSS is managed by the Astrophysical Research Consortium (ARC) for
the Participating Institutions. The Participating Institutions are the
American Museum of Natural History, Astrophysical Institute Potsdam,
University of Basel, University of Cambridge, Case Western Reserve
University, The University of Chicago, Drexel University, Fermilab,
the Institute for Advanced Study, the Japan Participation Group, The
Johns Hopkins University, the Joint Institute for Nuclear
Astrophysics, the Kavli Institute for Particle Astrophysics and
Cosmology, the Korean Scientist Group, the Chinese Academy of Sciences
(LAMOST), Los Alamos National Laboratory, the Max-Planck-Institute for
Astronomy (MPIA), the Max-Planck-Institute for Astrophysics (MPA), New
Mexico State University, Ohio State University, University of
Pittsburgh, University of Portsmouth, Princeton University, the United
States Naval Observatory, and the University of Washington.

\bibliography{ms}
\bibliographystyle{mn2e}

\begin{table}
  \caption{Radial Velocities}
  \label{tab-rvs}
  \begin{tabular}{@{}lcc}
    \hline
    {Date (TDB)} & {$v_1$ (km s$^{-1}$)} & {$v_2$ (km s$^{-1}$)} \\
    \hline
% forgot 32.184s correction for TAI->TDT
%  54021.37543 & $-55.49 \pm 2.75$ & $ 99.51 \pm 2.95$ \\
%  54021.40398 & $-57.00 \pm 2.89$ & $100.90 \pm 3.25$ \\
%  54081.21492 & $ 81.77 \pm 2.80$ & $-47.93 \pm 3.45$ \\
%  54106.23977 & $-41.12 \pm 3.06$ & $ 91.28 \pm 3.74$ \\
%  54106.27749 & $-46.12 \pm 3.12$ & $ 80.38 \pm 3.38$ \\
   54021.3758 & $-55.49 \pm 2.75$ & $ 99.51 \pm 2.95$ \\
   54021.4043 & $-57.00 \pm 2.89$ & $100.90 \pm 3.25$ \\
   54081.2152 & $ 81.77 \pm 2.80$ & $-47.93 \pm 3.45$ \\
   54106.2401 & $-41.12 \pm 3.06$ & $ 91.28 \pm 3.74$ \\
   54106.2778 & $-46.12 \pm 3.12$ & $ 80.38 \pm 3.38$ \\
  \hline
  \end{tabular}
  
  Barycentric radial velocities measured from the Keck data on Oct 13,
  2006, Dec 12, 2006 and Jan 6, 2007.  The errors include a systematic
  error of 1.75 km s$^{-1}$ which has been added in quadrature to each
  of the measured velocity uncertainties.  Dates are in Modified
  Julian Day (MJD) corrected to the solar system dynamic barycenter
  (TDB).
\end{table}

\begin{table*}
  \caption{Apparent brightness, colours, and limb-darkening of \name}
  \label{tab-pars1}
  \begin{tabular}{@{}lccccccc|cc}
    \hline
    {Filter} & {Star 1} & {Star 2} & {Total} & {$f_2/f_1$} & {$m_1-K_{\rm s,1}$} & {$m_2-K_{\rm s,2}$} & {$m_{tot}-K_{\rm s,tot}$} & {$u_1$} & {$u_2$} \\
    \hline
$u$&$ 21.95 \pm 0.05$ & $> 22.74^a$ &$ 21.9477 \pm 0.0461$ & $< 0.95^a$ &$  6.67^{+0.247}_{-0.323}$ & $> 7.2^a$ &$  7.3047 \pm 0.0462$ &0.713&0.734\\
$g$&$ 19.89^{+0.32}_{-0.23}$ &$ 21.26^{+1.22}_{-0.72}$ &$ 19.6159 \pm 0.0038$ &$  0.2833^{+0.43}_{-0.25}$ & $  4.61^{+0.072}_{-0.057}$ &$  5.73^{+0.875}_{-0.419}$ &$  4.9729 \pm 0.0043$ &0.829&0.814\\
$r$&$ 18.59^{+0.32}_{-0.23}$ &$ 19.70^{+1.06}_{-0.59}$ &$ 18.2573 \pm 0.0039$ &$  0.3581^{+0.46}_{-0.26}$ & $  3.31 \pm 0.07$ &$  4.18^{+0.670}_{-0.303}$ &$  3.6143 \pm 0.0043$ &0.808&0.777\\
$i$&$ 17.69^{+0.32}_{-0.23}$ &$ 18.66^{+0.91}_{-0.52}$ &$ 17.3164 \pm 0.0035$ &$  0.4100^{+0.48}_{-0.27}$ & $  2.41 \pm 0.07$ &$  3.13^{+0.505}_{-0.238}$ &$  2.6734 \pm 0.0041$ &0.696&0.672\\
$z$&$ 17.22^{+0.31}_{-0.23}$ &$ 17.99^{+0.69}_{-0.44}$ &$ 16.7874 \pm 0.0032$ &$  0.4943^{+0.50}_{-0.28}$ & $  1.94 \pm 0.06$ &$  2.46^{+0.298}_{-0.167}$ &$  2.1444 \pm 0.0038$ &0.611&0.589\\
$J$&$ 16.08^{+0.29}_{-0.22}$ &$ 16.47^{+0.42}_{-0.32}$ &$ 15.5019 \pm 0.0013$ &$  0.6979^{+0.53}_{-0.31}$ & $  0.80 \pm 0.05$ &$  0.94 \pm 0.06$ &$  0.8589 \pm 0.0024$ &0.481&0.428\\
$H$&$ 15.46^{+0.31}_{-0.23}$ &$ 15.74^{+0.41}_{-0.30}$ &$ 14.8373 \pm 0.0015$ &$  0.7672^{+0.58}_{-0.34}$ & $  0.18 \pm 0.05$ &$  0.22 \pm 0.06$ &$  0.1943 \pm 0.0025$ &0.453&0.398\\
$K_{\rm s}$&$ 15.28^{+0.31}_{-0.24}$ &$ 15.53^{+0.40}_{-0.30}$ &$ 14.6430 \pm 0.0020$ & $  0.7971^{+0.60}_{-0.36}$ & -- & -- & -- &0.377&0.329\\
    \hline
    \end{tabular}
  
  a : In the $u$ band the best-fit flux for the second star is zero, so we report 90\% limits
      on the magnitude and colours of this star.\\

  The apparent brightnesses of each component of \name, as well as the
  composite system brightness, as derived from our global fit.  We
  also list the adopted limb--darkening parameters $u_{1,2}$ for the
  components \citep{2000A&A...363.1081C,2004A&A...428.1001C}.  The
  total system brightness is better constrained than those of the
  individual components because of the latter's covariance with the
  radii of the stars, as well as the inclination of the system.  Due
  to the anticorrelation between the fluxes of the two stars, the flux
  ratio (column 5) has larger uncertainties.  The colours have smaller
  errors than the magnitudes since the fluxes in each band are
  strongly correlated for each star.  We quote a single error bar when
  the positive and negative uncertainties are the same to within 10\%.

\end{table*}

\begin{table*}
\caption{Binary Fit Parameters}
\label{tab-pars2}
\begin{tabular}{@{}lc}
\hline
{Parameter} & {Value} \\
\hline
$e$ cos($\omega$) & $0.00142 \pm 0.00068$ \\
$i$ (radians)     & $1.51445_{-0.00109}^{+0.00839} $ \\
$e$ sin($\omega$) & $-0.006 \pm 0.012$ \\
$T_{\rm 0}$ (TDB) & $52244.82052 \pm 0.00058$ \\
$P$ (days)        & $2.6390157 \pm 0.0000016$ \\
$R_1$($\rsun$)    & $0.639_{-0.090}^{+0.076} $ \\
$R_2$($\rsun$)    & $0.610_{-0.102}^{+0.083} $ \\
$M_1$($\msun$)    & $0.659 \pm 0.031$ \\
$M_2$($\msun$)    & $0.619 \pm 0.028$ \\
$\gamma$ (km s$^{-1}$) & $19.09 \pm 1.28$ \\
\hline
\hline
$M_1/R_1$($M_\odot/R_\odot$) & $  1.032^{+0.171}_{-0.112}$ \\
$M_2/R_2$($M_\odot/R_\odot$) & $  1.016^{+0.204}_{-0.124}$ \\
$R_1+R_2$ ($\rsun$)          & $  1.248^{+0.021}_{-0.048}$ \\
$R_2/R_1$                    & $  0.955^{+0.304}_{-0.244}$ \\
$a/R_1  $                    & $ 13.652^{+2.220}_{-1.455}$ \\
$a/R_2  $                    & $ 14.299^{+2.850}_{-1.707}$ \\
$K_1$ (km s$^{-1}$)          & $  80.896\pm 1.811$ \\
$K_2$ (km s$^{-1}$)          & $  76.062\pm 4.204$ \\
$K  $ (km s$^{-1}$)          & $ 156.959\pm 5.901$ \\
$log(g_1$ [cm s$^{-2}$])     & $  4.646^{+0.131}_{-0.098}$ \\
$log(g_2$ [cm s$^{-2}$])     & $  4.659^{+0.158}_{-0.111}$ \\
$\rho_1$  (g cm$^{-3}$)      & $  3.563^{+2.032}_{-1.023}$ \\
$\rho_2$  (g cm$^{-3}$)      & $  3.849^{+2.790}_{-1.223}$ \\
$M_1+M_2$ ($M_\odot$)        & $   1.278\pm 0.054$ \\
$M_2/M_1$                    & $   0.940\pm 0.034$ \\
$a$ ($\rsun$)                & $   8.718\pm 0.123$ \\
$a$ (AU)                     & $   0.041\pm 0.001$ \\
$b$ ($R_\odot$)              & $  0.396^{+0.005}_{-0.050}$ \\
$T_2/T_1$                    & $  0.947^{+.032}_{.016}$\\
\hline
\end{tabular}

Best fit system and physical parameters of \name\ (above line) and
derived parameters (below line).  The fitting process is described in
Section~\ref{sec-model}.  Uncertainties in the parameters are derived
from the distribution of best-fit values for the 10$^5$ synthetic
light curves.  We quote a single error bar when the positive and
negative uncertainties are the same to within 10\%.

\end{table*}

% If we decide to include all the data...
% \include{table1b}

\label{lastpage}

\end{document}